\documentclass[runningheads]{llncs}
\usepackage[T1]{fontenc}

\usepackage{xcolor}
\usepackage{graphicx}

\usepackage{comment}
\usepackage{booktabs}

\let\classvec\vec  % save old \vec
\let\vec\relax
\usepackage{amsmath}
\usepackage{amssymb}
\usepackage{amsfonts}
\let\vec\classvec

\usepackage{multirow}
\usepackage{graphicx} 

\usepackage[hyphens]{url} 
\PassOptionsToPackage{hyphens}{url}
\usepackage{hyperref}
\usepackage{subcaption}

\begin{document}
\title{Privacy-Preserving Iris Recognition: \\ Performance Challenges and Outlook}

\author{Christina Karakosta\inst{} \and
Lian Alhedaithy\inst{} \and
William J. Knottenbelt\inst{} }
\authorrunning{C. Karakosta et al.}

\institute{Department of Computing, Imperial College London\\
\email{\{c.karakosta20,lian.alhedaithy24,w.knottenbelt\}@imperial.ac.uk}}
\maketitle             
\begin{abstract}
Iris-based biometric identification is increasingly recognized for its significant accuracy and long-term stability compared to other biometric modalities such as fingerprints or facial features. However, all biometric modalities are highly sensitive data that raise serious privacy and security concerns, particularly in decentralized and untrusted environments. While Fully Homomorphic Encryption (FHE) has emerged as a promising solution for protecting sensitive data during computation, existing privacy-preserving iris recognition systems face significant performance limitations that hinder their practical deployment. This paper investigates the performance challenges of the current landscape of privacy-preserving iris recognition systems using FHE. Based on these insights, we outline a scalable privacy-preserving framework that aligns with all the requirements specified in the ISO/IEC 24745 standard. 
Leveraging the \emph{Open Iris} library, our approach starts with robust iris segmentation, followed by normalization and feature extraction using Gabor filters to generate iris codes. We then apply binary masking to filter out unreliable regions and perform matching using Hamming distance on encrypted iris codes. The accuracy and performance of our proposed privacy-preserving framework is evaluated on the CASIA-Iris-Thousand dataset. Results show that our privacy-preserving framework yields very similar accuracy to the cleartext equivalent, but a much higher computational overhead with respect to pairwise iris template comparisons, of $\sim 120\,000 \times$. This points towards the need for the deployment of two-level schemes in the context of scalable \textit{1--N} template comparisons.

\end{abstract}
\section{Introduction}
\label{sec:introduction}
Biometric identification systems have become one of the most reliable forms of personal authentication in the digital world. In particular, iris-based identification has emerged as highly reliable due to the uniqueness and stability of iris patterns throughout an individual's lifetime~\cite{Amour-2020}. Studies indicate that an iris code can contain over 200 bits of entropy, which is significantly more than fingerprints~\cite{Young-2013}, providing enough discriminatory power to distinguish over 8 billion people~\cite{Daugman-collision}. This high entropy makes iris recognition more accurate than facial recognition.  

In the early 1990s, John Daugman pioneered the foundational work in iris recognition by developing algorithms for iris pattern encoding and matching with very high accuracy. Daugman's approach encodes the iris texture using multi-scale 2D Gabor wavelets and compares the binary iris codes using the fractional Hamming distance as a similarity metric~\cite{Daugman}. Highly reliable identification is possible by testing for statistical independence between iris codes, achieving rapid matching via XOR operations on IrisCodes. However, biometric data are highly sensitive Personally Identifiable Information (PII). This creates serious security and privacy threats in the event of a data breach. A stolen biometric template could be reused by attackers for identity theft or cross-matching across databases. If an attacker accesses a biometric template from one system, that individual’s identity could be compromised in all other systems using the same biometric. Such risks are increased in decentralized environments where data and computation are distributed across multiple entities without a central authority. In response to these concerns, international standards such as ISO/IEC 24745~\cite{ICO/IEC_24745} provide requirements and guidelines for protecting biometric information, emphasizing principles such as irreversibility, unlinkability, and confidentiality of biometric data throughout its lifecycle. 

Therefore, there is a critical need to handle biometric data differently than passwords, in order to prevent reconstruction, cross-linking, and unauthorized use of these immutable personal identifiers. Privacy-Enhancing Technologies (PETs) have been explored to address these challenges. Various approaches have been proposed for privacy-preserving biometric authentication, including Fully Homomorphic Encryption (FHE), Secure Multi-Party Computation (MPC), and other hybrid approaches. 

\textbf{Problem Statement:} However, existing iris recognition systems using PETs, especially FHE-based solutions, face serious scalability constraints. There is a fundamental trade-off between privacy, performance, and usability. Current fully homomorphic approaches incur substantial computation and communication overhead when scaling to large databases or real-time identification scenarios. One analysis showed that transmitting a single 12\,800-bit iris code under typical FHE would require about 37 MB of ciphertext~\cite{taceo}. Storing an encrypted database of 100\,000 iris codes could require 3.5 terabytes of storage space~\cite{taceo}, which is an impractical burden on network and storage resources. Similarly, computation latency is a challenge, since executing a homomorphic iris code comparison can take seconds of CPU time~\cite{Song-2020}, and performing thousands or millions of such comparisons for identification would be prohibitively slow without significant optimization or parallelization. 

Another bottleneck is the handling of rotation alignment and masking of iris codes under encryption. Iris templates must be aligned (rotationally shifted) according to the eye orientation differences, which normally requires multiple XOR comparisons of shifted codes. Doing this over encrypted data is extremely expensive, involving costly cyclic shifts on ciphertext or re-encrypting multiple shifted copies. Bassit et al.’s study confirms that supporting rotations under encryption reduces matching performance, whereas alignment-invariant methods (such as Bloom filter representations) maintain a speed advantage~\cite{Bassit-2020}. Thus, the key barriers to practical deployment are: \textit{(i)} the computational cost of homomorphic operations on large binary iris codes, \textit{(ii)} the complexity of template alignment and masking when data is encrypted, and \textit{(iii)} the explosion of ciphertext size which affects storage and bandwidth.

This research aims to survey these bottlenecks in existing FHE-based iris recognition frameworks and then outline an end-to-end privacy-preserving iris biometric system suitable for real-world use. We identify how current methods lack a balance between privacy and performance, and propose architectural and algorithmic optimizations to bridge that gap. In particular, we present a framework for trustworthy, privacy-by-design digital identity systems using FHE that addresses the above challenges. Our approach incorporates efficient iris code generation and matching techniques for encrypted operations, including methods to deal with masks and rotations in a cost-effective fashion. We also ensure the framework aligns with international standards (ISO/IEC 24745) for biometric information protection, embedding privacy and security requirements by design.

\textbf{Contributions:} The main contributions of this paper are:
\vspace{-7mm}
\begin{itemize}
    \item \emph{Performance Analysis:} We provide a structured survey of prior privacy-preserving iris recognition systems and analyze their performance limitations. We highlight performance bottlenecks in encrypted matching, mask handling, rotation, storage, and communication overheads.
    \item \emph{Proposed Framework:} We integrate a standard iris-recognition pipeline (segmentation, normalization, Gabor encoding, masking) into an FHE-based matching workflow. We evaluate the masking and rotation cost, and the feasibility of performing end-to-end iris-recognition under fully homomorphic operations.
    \item \emph{Implementation and Evaluation:} We implement a proof-of-concept system using a lattice-based FHE scheme and evaluate cleartext vs.\ encrypted matching on the CASIA-Iris-Thousand dataset. We quantify computational latency, communication overhead, ciphertext size, and the impact of parameter choices on performance.
\end{itemize}

The remainder of this paper is organized as follows. \textbf{Section~\ref{sec:background}} provides background on biometric privacy challenges and an overview of privacy-enhancing technologies applicable to biometrics. \textbf{Section~\ref{sec:related_work}} reviews related work in privacy-preserving iris recognition and compares different approaches. \textbf{Section~\ref{sec:system_architecture}} describes the proposed system architecture, participants, and security model. \textbf{Section~\ref{sec:methodology}} details our methodology, including the iris code generation process and the homomorphic encryption and matching techniques. \textbf{Section~\ref{sec:evaluation}} presents the experimental setup and metrics for evaluation. \textbf{Section~\ref{sec:discussion}} discusses the results, performance analysis, and insights from the evaluation. Finally, \textbf{Section~\ref{sec:conclusions}} concludes the paper and outlines directions for future work.

\section{Background}
\label{sec:background}
\subsection{Biometric Privacy Concerns}
Biometric data requires different handling than conventional authentication secrets, like passwords or tokens, because of its permanence and link to an individual’s identity~\cite{Gupta-2016}. Biometrics such as fingerprints, face images, or iris patterns are uniquely linked to a person and effectively irreplaceable. 
\begin{itemize}
    \item \textbf{Irreversible Disclosure:} A stolen biometric template can be used to reconstruct or impersonate the original trait. Researchers have demonstrated that compromised iris code data could potentially be used to generate a fake iris image. Thus, biometric identifiers must be highly protected to prevent irreversible privacy violation.
    \item \textbf{Cross-Context Linkage:} Because biometric identifiers are globally unique to a person, leaks can enable linkage across databases. For example, if one’s iris code from an authentication system is exposed, an adversary could scan other databases or systems for a matching code to link that person’s activities across different services. This could lead to a violation of unlinkability, mass surveillance or profiling.
    \item \textbf{Identity Theft and Fraud:} Stolen biometrics can be used for identity theft or unauthorized access. Attackers might use fingerprints or high-resolution iris images to bypass biometric access control (presentation attacks). Alternatively, an attacker could create a synthetic identity by combining someone’s biometric with false demographic data. Since biometrics are often considered highly secure, many systems may not have secondary verification, making a stolen biometric a single point of failure for identity.
    \item \textbf{Sybil Attacks in Digital ID:} In decentralized identity systems, a key threat is a Sybil attack, where one person tries to obtain multiple identities. Biometrics are often employed to resist Sybil attacks by enforcing uniqueness ($one\ person = one\ ID$). However, if biometric protections fail, an attacker could duplicate or forge biometrics to create many fake identities. For instance, without robust protection, a malicious user might replay an iris code to register multiple times under different names. Preventing Sybil attacks requires ensuring the biometric cannot be easily duplicated or abused across the system.
\end{itemize}

In summary, biometrics offer strong authentication benefits due to their uniqueness, but these same properties demand stronger safeguards. Reconstruction, linkage, identity fraud, and Sybil attacks are primary threats. This necessitates specialized privacy protections (e.g.,\ template encryption, secure storage, liveness detection) beyond what is needed for traditional secrets.

\subsection{Privacy-Enhancing Technologies for Biometrics}
Privacy-Enhancing Technologies (PETs) can secure biometric data even during processing. Key PETs approaches in the biometric field include the following:
\begin{itemize}
    \item \textbf{Fully Homomorphic Encryption (FHE):} FHE allows computations to be performed on encrypted data without ever decrypting it. In a biometric system, this means an iris or fingerprint template can remain encrypted while the server computes a match score. FHE guarantees confidentiality because the server, or any eavesdropper, sees only encrypted results. This directly addresses privacy and confidentiality requirements (ISO/IEC 24745 emphasizes confidentiality and irreversibility). FHE performance is limited by homomorphic operations that are orders of magnitude slower than normal operations and produce much larger ciphertexts. For example, basic FHE schemes might encrypt each bit of an iris code into a large ciphertext polynomial, leading to huge data sizes and computational overhead for each comparison. FHE also assumes an ``honest-but-curious'' threat model where the server will not alter computations maliciously. Despite these costs, FHE provides a trustable model where even an untrusted server cannot learn biometric data, so a single server suffices, simplifying deployment if the performance can be managed.
    \item \textbf{Secure Multi-Party Computation (MPC):} MPC protocols distribute the computation across multiple parties so that no single party can see the entire input. In a biometric setting, the biometric template can be split into shares held by different servers. The servers then engage in an interactive protocol to jointly compute a function, without revealing their shares. The strength of MPC is that it can often be more efficient than FHE for certain operations by leveraging lightweight secret-sharing schemes and parallel computation. For instance, Worldcoin’s deployment uses an MPC protocol to compare iris codes for uniqueness across millions of users~\cite{Worldcoin}. The iris code and its mask are split into shares and distributed to multiple servers, which perform comparison without reconstructing the code~\cite{Worldcoin}. Many linear operations (XORs, additions) can be done with no communication in MPC, and clever protocols exist for non-linear operations (e.g.,\ counting bits) with far less communication than naive methods~\cite{taceo}. MPC relies on a trust model where a threshold of parties must remain honest and non-colluding. While it introduces network latency due to message exchanges for every operation, MPC has demonstrated practical scalability.
    \item \textbf{Functional Encryption (FE):} Functional Encryption is a newer approach where possessing a certain secret key allows one to learn only a specific function of the encrypted data. In biometrics, FE enables a server to use a specific key that reveals only the match score or a binary decision from encrypted data. For example, a policy could be encoded so that decryption reveals ``\textit{match if $distance < threshold$}'' without exposing raw biometric templates. In this way it can enforce access control at a data level, while the server accesses only a permitted result and never the raw biometric. FE can be thought of as an extension of attribute-based encryption for computations. However, general functional encryption is complex and often less efficient. While prototypes like inner product schemes exist~\cite{Ernst-2023}, FE is currently not efficient enough for high-dimensional biometrics and may still require trust in a key issuer or authority for key distribution and policy enforcement.
    \item \textbf{Hybrid Approaches:} Bloom filters provide a cancelable, privacy-protecting transformation for biometric templates. They encode feature positions into bit strings, obscuring exact bits while allowing similarity comparisons (e.g.,\ weighted Hamming distance)~\cite{Bassit-2020}. Bloom filters are fast, space-efficient, potentially non-invertible, and inherently handle rotations without brute-force shifting. However, poor designs risk linkability~\cite{Patgiri-2021}. Hybrid approaches, like Bassit et al.~\cite{Bassit-2020}, address this by combining randomized Bloom filters with homomorphic encryption. This achieves the speed of Bloom filters with the security of encryption, significantly outperforming pure FHE while meeting ISO privacy standards. While promising, hybrid schemes add system complexity and may offer weaker security guarantees than FHE, as protection relies partially on transformation rather than encryption.
    \item \textbf{Trusted Execution Environments (TEEs):} Trusted Execution Environments (TEEs), such as \textit{Intel SGX} or \textit{ARM TrustZone}, provide hardware-isolated execution for sensitive code and can perform iris matching on plaintext data inside an enclave. This eliminates the heavy computational overhead of FHE since Hamming distance can be computed directly on the plaintext template. However, their security relies on trusting hardware vendors, firmware updates, and resistance to side-channel attacks~\cite{Georgia-2025,Munoz-2023,Nilsson-2020,VanBulck-2019,Wang-2017}. Thus, the reliance on TEEs introduces security risks, reinforcing the decision of building a cryptographic and hardware-agnostic solution.
\end{itemize}

Each of these approaches can be part of a privacy-by-design biometric system. In practice, combinations may be used to leverage the advantages of each.

\subsection{Biometric Template Protection Mechanisms}
Beyond cryptographic PETs, several mechanisms protect biometric templates against irreversible disclosure and linkage attacks. \emph{Cancelable biometrics} use non-invertible transformations (e.g.,\ random projection, BioHashing) that allow revocation and re-issuance if a template is compromised. \emph{Secure sketch and fuzzy commitment schemes} cryptographically bind biometric data to auxiliary helper data without leaking the underlying template. Even if stored data are leaked, the original iris code cannot be reconstructed. Finally, \emph{fuzzy extractors} derive stable cryptographic keys from noisy biometric signals while exposing only public helper data with provably limited leakage. Strengthened fuzzy extractors increase entropy extraction and reduce cross-database linkability.

Our work focuses on protecting the matching process using FHE, and such template-protection methods could be layered on top of our design.

\section{Related Work}
\label{sec:related_work}
Early research on privacy-preserving biometric recognition explored both cryptographic protocols and signal transformations. In the context of iris recognition, some of the first solutions used partial homomorphic encryption or two-party protocols to compute the Hamming distance securely. Blanton and Gasti (2011)~\cite{Blanton-2011} proposed secure protocols for iris and fingerprint identification using an additively homomorphic cryptosystem (DGK) and garbled circuits~\cite{Damgard-2008}. Their protocol could compare 2048-bit iris codes in about 150 ms, highlighting the feasibility of fast encrypted matching~\cite{Blanton-2011}. However, the efficiency came with a trade-off: the smaller ciphertext size in their scheme meant lower cryptographic security, underscoring the tension between efficiency and strength. Subsequent works moved to stronger encryption at the cost of speed. Kulkarni et al.\ implemented an iris authentication using a somewhat homomorphic encryption (SHE) scheme, achieving 58 seconds server time for matching a 2\,048-bit iris code, with a few rounds of interaction~\cite{Kulkarni-2013}. Karabat et al.\ (2015) proposed a threshold homomorphic encryption protocol achieving server and client computation times of $\sim$6.1 s and 2.1 s, respectively~\cite{Karabat-2015}. These studies highlight that FHE, while more secure, requires significant optimization for practicality.

More recent works have focused on leveraging improved FHE schemes and data handling techniques. Morampudi et al.\ (2020) introduced a privacy-\hspace{0pt}preserving iris authentication method based on the BFV homomorphic encryption scheme~\cite{Morampudi-2020}. They addressed one of the major performance issues, rotation alignment, by generating rotation-invariant iris codes so that the costly cyclic shifts under encryption could be avoided. They also used SIMD batching on ciphertexts to pack multiple iris code bits into a single polynomial, amortizing operation costs. Their system satisfied the ISO/IEC 24745 requirements for biometric template protection and reported no loss in accuracy compared to an unencrypted system. In fact, they achieved an Equal Error Rate (EER) of 0.19\% on the standard CASIA-Iris V1 dataset, which is on par with cleartext iris recognition performance. Their homomorphic Hamming distance for a 2\,560-bit iris code took only about 0.0185 seconds. 

On the MPC side, Worldcoin’s World ID system represents a state-of-the-art deployment for secure and scalable iris-based identification. Rather than performing heavy cryptography on a single server, Worldcoin secret-shares each iris code among multiple nodes and uses MPC to check for duplicates~\cite{Worldcoin-whitepaper}. Their protocol enables private, million-scale matching via efficient distributed Hamming distance and threshold checks~\cite{taceo}. A key technical innovation is the use of an ``efficient dot product'' method to compute Hamming distances with communication costs independent of vector size~\cite{taceo}. By converting binary shares to arithmetic shares over a ring (e.g.,\ 16-bit integers), the protocol reduces the complexity of a 12\,800-bit Hamming distance computation from 25\,000 AND gates to a fixed constant. This efficiency enables significant scalability, considering that  sending a single iris code of 32 MB FHE ciphertext would exceed the communication cost of their entire MPC protocol for a 100\,000 record database. Thus, while MPC outperforms FHE in multi-server environments, it necessitates higher operational complexity regarding server synchronization and strictly non-colluding parties.

Template transformations, such as cancelable biometrics and fuzzy schemes, complement cryptographic methods. While less robust than FHE or MPC, these techniques prevent reconstruction if data is stolen. Yang et al.~\cite{Yang-2023} highlight the benefit of hybrid approaches, where encrypting transformed templates adds defense-in-depth against decryption failures. However, challenges remain regarding biometric variability (e.g.,\ rotation, noise) and standards compliance.

Scalability and latency remain primary bottlenecks for secure biometrics. FHE struggles with scaling, while MPC requires multiple servers and heavy communication. Transformations like Bloom filters offer speed but demand rigorous security analysis. Building on~\cite{Morampudi-2020,Worldcoin}, our work targets optimized performance deployment by strictly addressing encrypted iris rotations and masking, which were often simplified or brute-forced in prior works. Our design ensures unlinkability and prevents raw biometric exposure in compliance with privacy standards.

\section{System Architecture and Security Model}
\label{sec:system_architecture}
\begin{figure}
\includegraphics[width=\textwidth]{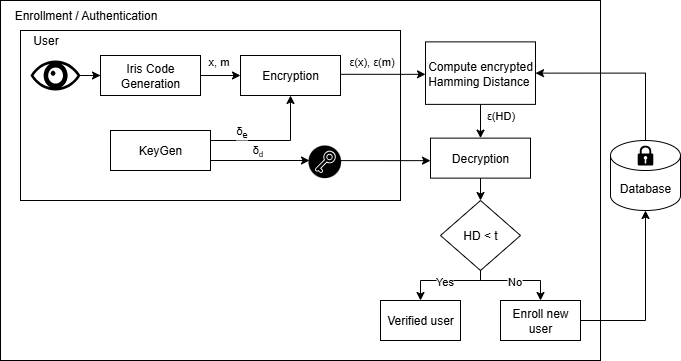}
\caption{FHE iris recognition system} \label{fig1}
\end{figure}

A high-level architecture of the proposed privacy-preserving iris recognition system is presented in Figure~\ref{fig1}. The user-side device captures the iris image and locally computes the iris code $(x)$ and mask $(m)$. The iris code bits and masks are then encrypted $(\varepsilon(x),\varepsilon(m))$ using the user’s public key ($\delta_\varepsilon$) before being sent to the server. The encrypted iris code $(\varepsilon(x))$ is used as a query for matching against existing encrypted templates. The server (untrusted) performs homomorphic computations on encrypted data to get the Hamming Distance result $(\varepsilon(HD))$. Then, the result $(\varepsilon(HD))$ is decrypted using the private decryption key ($\delta_d$), and compares the decrypted result with a threshold  $(t)$. If the Hamming Distance is below the threshold the user is already verified, otherwise the system enrolls the new user in the database. 

Participants in our system include: \textit{(1) The User}, who owns their biometric data. This could be an individual enrolling their iris on a personal device or an orb-like device. This device is responsible for the iris code generation and encryption of the iris code with the user’s public key. We assume the user keeps their FHE private key secure, which is needed to decrypt match results. \textit{(2) The Server}, which is untrusted, stores encrypted biometric templates and performs homomorphic computations. The server is considered ``\emph{honest-but-curious}'' by computing the Hamming Distance honestly, but may attempt to infer sensitive information from any data it sees. In our design, the server only handles encrypted data, and we do not require any trusted hardware on the server side. In an identification scenario (\textit{1--N} matching), the server might play the dual role of providing the result to the user or an authorized verifier. However, our current implementation evaluates only \textit{1--1} individual pairwise comparisons. Extending the system to encrypted \textit{1--N} identification remains future work.

\textbf{Threat Model:} Our threat model assumes that all the main infrastructure, including the network and the server, is untrusted. Communications pass over an untrusted network, an eavesdropper could intercept messages, and an active attacker might attempt ``\emph{man-in-the-middle}'' modifications. Thus, we use standard network security (TLS) and ensure that all biometric data are encrypted. The server is considered curious, but non-colluding. It will follow the protocol steps, but is curious to learn sensitive information. It may also be compromised by an adversary who gains read-access to stored data. The user side is assumed to be trusted with regard to their own biometric. The user will not intentionally leak their private key or biometric data. However, malware on a user’s device is a potential threat. If an attacker steals the user’s private key, they could decrypt the user’s template or results. Regarding key storage, we specifically avoid centralized management. While traditional systems secure master keys inside organization trust boundaries using Hardware Security Modules (HSMs), reported hacks on HSMs~\cite{Ledger-2019} demonstrate that even FIPS-certified hardware can be compromised. Thus, key protection on the client side, via secure storage or use of a hardware token, is assumed. Also, multiple users do not collude by sharing keys or data, since each user has their own key pair.

\emph{Insider threats} with database access may attempt to inspect ciphertexts or misuse system metadata. Our design ensures that ciphertexts leak no biometric information, but we assume secure management of private keys. \emph{Malicious server outputs} might return incorrect encrypted scores or manipulate threshold decisions. Detecting such behavior requires verifiable computation or redundant computation across multiple servers, which we discuss as future work. \emph{Key loss} of the user's private key requires re-enrolled under a fresh key. This prevents cross-database linkage but requires secure user-side key backup.

Our security goals are: \textit{(1) No Leakage of Raw Biometric Data:} The server or any external adversary should not learn any information about the user’s iris image or iris code beyond what is revealed by the final match result. Even partial leakage, like number of bits matching, should be minimized or provably bounded. \textit{(2) Encrypted Matching:} All biometric comparisons for matching are calculated on encrypted data. The system never needs to decrypt templates on the server side. \textit{(3) Template Protection (Irreversibility):} The stored encrypted template should not allow an attacker to reconstruct the original iris. Even if an attacker somehow obtained a user’s encrypted iris code and also knew the encryption scheme, without the key they should not be able to recover the code or any significant part of it. Our use of FHE guarantees this under standard cryptographic assumptions. \textit{(4) Unlinkability:} It should be infeasible for an attacker to determine if two encrypted templates belong to the same person (unless they are an authorized party performing a match). This means, for example, if a user enrolls in two different databases with the same iris, the encrypted templates should be unrelated. We achieve this by having each user use their own key so that encryptions of the same iris under different keys cannot be linked.

In summary, our architecture addresses threats from network adversaries,
curious or partially compromised servers, insider misuse, and client-side key compromise, while ensuring privacy through end-to-end encryption processing.

\section{Methodology}
\label{sec:methodology}
\begin{figure}
\includegraphics[width=\textwidth]{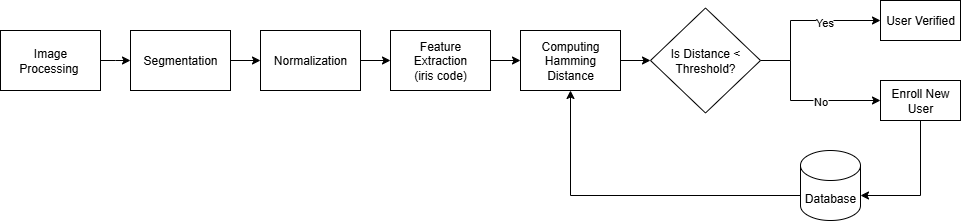}
\caption{Iris code feature extraction and matching process} \label{fig2}
\end{figure}

Our processing pipeline consists of the following stages of iris recognition: \textit{(1) segmentation}, \textit{(2) normalization}, \textit{(3) feature extraction}, and \textit{(4) matching}. Figure~\ref{fig2} outlines these stages in sequence. 

\vspace{-0.6cm}
\vspace{-1.25cm}

\subsection{Iris Code Generation}
\paragraph{Preprocessing:} Input eye images are already in grayscale. We assess image quality because poor-quality images can degrade recognition and also waste computational resources if processed under encryption. We detect and remove specular highlights (reflections) which appear as saturated white spots on the iris. A simple method is using percentile thresholding and then inpainting or smoothing over them, since specularities can falsely appear as unique features if not handled. After reflection removal, the image is enhanced using contrast normalization to improve iris feature detection in the next steps.  

\paragraph{Segmentation:} 
This step isolates the iris region from the rest of the eye image by determining the iris inner (pupil) boundary and outer (sclera) boundary, and identifying occlusions by eyelids or eyelashes. We employ a deep learning-based segmentation approach using the \textit{Open Iris} framework~\cite{open-iris-repo}, which utilizes a convolutional neural network (CNN) trained on diverse iris datasets to achieve robust boundary detection. The model outputs precise pixel-level segmentation masks that distinguish the iris region from the pupil, sclera, eyelids, and eyelashes. This modern approach provides superior accuracy compared to classical methods, particularly for handling challenging cases with poor lighting, partial occlusions, or off-angle captures. The output of segmentation is a binary mask indicating which pixels belong to the clear iris region ($1=\text{valid iris}$, $0=\text{occluded/invalid}$), along with the detected inner and outer iris boundaries.

\paragraph{Normalization:} 
Once the iris region is segmented, we transform it to a normalized coordinate system to account for differences in iris size, pupil dilation, and imaging angle. We use the standard Daugman rubber-sheet model~\cite{Daugman}: the annular iris region is mapped to a fixed-size rectangle with angular coordinate $\theta$ along the horizontal axis and radial coordinate $r$ along the vertical axis. In our implementation, we adopt a resolution of 512 angular samples and 64 radial samples, yielding a normalized image of size $512 \times 64$. This effectively "unrolls" the iris into a rectangular band where each row corresponds to a concentric circular ring of the iris. The normalization compensates for iris size differences by stretching or compressing radii to a constant length, and accounts for pupil dilation by anchoring the mapping between the pupil and iris boundaries.

\paragraph{Feature Extraction:} 
The normalized iris texture is encoded into a binary iris code using multi-scale 2D Gabor wavelets to extract local phase information~\cite{Daugman}. We apply Gabor filters at multiple scales (wavelengths of 16 and 32 pixels) to capture texture features at different resolutions. Each Gabor filter produces a complex-valued response, and we quantize the phase into two bits using quadrant encoding: the real and imaginary components determine which of four quadrants the phase vector falls into, yielding two binary values per filter response. This process is applied across the entire normalized iris image, generating two binary matrices (one for each scale). These matrices are then flattened and concatenated to form the final iris code. In our implementation, with a $128 \times 512$ effective sampling grid (after downsampling from the $512 \times 64$ normalized image) and two scales, we generate an iris code of $128 \times 512 \times 2 = 131\,072$ bits. However, for computational efficiency in the FHE matching stage, we downsample this to a code length of $32 \times 512 = 16\,384$ bits, which provides a balance between discriminative power and encryption overhead.

\paragraph{Mask Generation:} 
Alongside the iris code, we generate a binary mask of the same dimensions indicating which bits of the iris code are valid. Bits are marked as invalid (0 in the mask) due to occlusion by eyelids or eyelashes, specular reflections, or regions near segmentation boundaries where texture reliability is low. The segmentation stage identifies these regions, and we propagate this information through the normalization and feature extraction stages to mark the corresponding code bits as masked. Masking is crucial for robust matching, as it prevents noise from contributing to the Hamming distance. In our system, the mask is encrypted alongside the iris code to enable privacy-preserving masked matching. The mask and code together form the biometric template.

\paragraph{Code Alignment:} 
Before encryption, we prepare for handling rotation misalignment between enrollment and verification captures. Eye rotation or head tilt can cause the iris code to shift circularly along the angular dimension. To handle this, during the matching phase we perform a rotational search by computing Hamming distances across multiple circular shifts of the iris code. In our implementation, we allow a search window of $\pm 15$ columns (out of 512 angular samples), which corresponds to approximately $\pm 10.5^\circ$ of physical rotation. This range is sufficient to handle natural head tilt variations in cooperative acquisition scenarios typical of datasets like CASIA-Iris-Thousand. Under FHE, this rotational search can be implemented efficiently using ciphertext rotation operations (automorphisms in the BFV scheme)~\cite{Song-2020}, though in our current implementation we perform the search by encrypting and comparing multiple shifted versions.

\paragraph{}By the end of these steps, the user has an iris code and a corresponding mask (both binary vectors). At this point, both are still in cleartext on the user side. The next phase involves encrypting these templates and transmitting them to the server for privacy-preserving matching.

\subsection{Encryption and Hamming Distance Computation}
\paragraph{Fully Homomorphic Encryption Scheme:} We employ a Fully Homomorphic Encryption technique based on the General Learning with Errors (GLWE) scheme, implemented through Vaultree's \texttt{venumML}~\cite{vaultree2025} library. The GLWE scheme provides the cryptographic foundation for all encrypted computations, enabling privacy-preserving iris template matching without data exposure. Other libraries such as Microsoft SEAL~\cite{SEAL-2017} and OpenFHE~\cite{OpenFHE-2022} provide highly optimized BFV / BGV implementations and efficient bitwise operations, which would likely provide lower latency than a full PPML stack. Similarly, HElib~\cite{HElib-2014} offers efficient ciphertext packing (SIMD), making it suitable for \textit{1--N} matching scenarios.

The General Learning With Errors (GLWE) scheme is a lattice-based cryptographic primitive that generalizes the Learning with Errors (LWE) problem. It introduces structured noise over polynomial rings, enabling compact ciphertexts and efficient homomorphic operations. GLWE is considered secure against classical and quantum attacks, making it suitable for post-quantum cryptography.

Given a binary iris code vector and a corresponding mask, both arrays are flattened and each element is encrypted as follows:
\vspace{-0.2cm}
\begin{align}
    c_i = \texttt{ctx.encrypt}(b_i), \quad b_i \in \{0,1\}
\end{align}
\vspace{-0.5mm}
where \texttt{ctx} is the encryption context initialized using \texttt{SecretContext()}.

\paragraph{Encrypted Matching with Hamming Distance:}
The template matching is performed using normalized Hamming distance calculation between iris codes. The fractional Hamming distance is computed using the generated masks to exclude unreliable bits from the comparison. The fractional Hamming distance (HD) quantifies the dissimilarities between two iris‐codes, $\mathbf{code}_A$ and $\mathbf{code}_B$, using their corresponding mask vectors, $\mathbf{mask}_A$ and $\mathbf{mask}_B$, as follows: 
\vspace{-0.2cm}
\begin{align}
    HD=\frac{||\mathbf{code}_A \oplus \mathbf{code}_B \cap\ \mathbf{mask}_A \cap \mathbf{mask}_B||} {||\mathbf{mask}_A \cap \mathbf{mask}_B||}
\end{align}
\vspace{-0.3cm}

In our system, the Hamming Distance (HD) is computed between encrypted iris codes homomorphically:
\begin{itemize}
    \item The encrypted XOR of two bits $a$ and $b$ is computed as:
    \vspace{-0.3cm}
    \[
    \text{XOR}(a, b) = a + b - 2ab
    \]
    \vspace{-0.8cm}
    \item The valid mask is computed as the element-wise product of the two encrypted masks.
    \item The masked XOR values are summed to count the differing bits.
    \item The total number of valid bits is summed homomorphically.
\end{itemize}

Let $D$ denote the encrypted count of differing bits, and $N$ the count of valid comparisons. The encrypted normalized Hamming distance is $D/N$.

After computation, the encrypted outputs $D$ and $N$ are sent back to the user, who decrypts them using their private key. The final Hamming distance is computed in cleartext:
\[
\text{HD} = \frac{D}{N}
% \text{HD} = D/N
\]
Then, the HD is compared with a threshold ($threshold = 0.35$) to determine whether the two iris templates match.

\section{Experimental Evaluation and Results}
\label{sec:evaluation}
We evaluated our framework on a GNU/Linux system with an NVIDIA GeForce GTX 1\,080 GPU. We utilized the \textit{Open Iris} library~\cite{open-iris-repo} for biometric processing and Vaultree’s \textit{venumML}~\cite{vaultree2025} library for all FHE operations, configured for a 128-bit security level. The analysis was performed on the CASIA-Iris-Thousand dataset~\cite{CASIA-Iris}, which contains 20\,000 iris images from 1\,000 subjects (10 images for each eye of 1\,000 subjects). After removing images with glasses, as they affect the image quality, we created four subsets of varying sizes comprising 50, 100, 500 and 1\,000 subjects. Then, we performed all possible intra-class (genuine) and inter-class (imposter) comparisons to establish a cleartext performance baseline and assess scalability. Further details on the experimental setup are provided in Appendix~\ref{appendix:a1}.
\vspace{-2mm} 
\begin{itemize}
    \item \textbf{CASIA 50}: A database of 50 subjects containing 755 iris templates.
    \item \textbf{CASIA 100}: A database of 100 subjects containing 1\,487 iris templates.
    \item \textbf{CASIA 500}: A database of 500 subjects containing 6\,777 iris templates.
    \item \textbf{CASIA 1K}: A database of 1,000 subjects containing 14\,191  iris templates.
\end{itemize}

\subsection{Cleartext Performance and Scalability Analysis}
The cleartext matching performance demonstrated high accuracy and stability as the database size increased, as summarized in Table~\ref{tab:scalability_results}. The Equal Error Rate (EER) remained consistently low, ranging between 0.12\% and 0.27\%, while the F1 Scores exceeded 0.99 across all subsets, confirming the robustness of the feature extraction and matching algorithm. 

\begin{figure}[h]
    \centering
    \begin{subfigure}[b]{0.46\linewidth}
        \includegraphics[width=\linewidth]{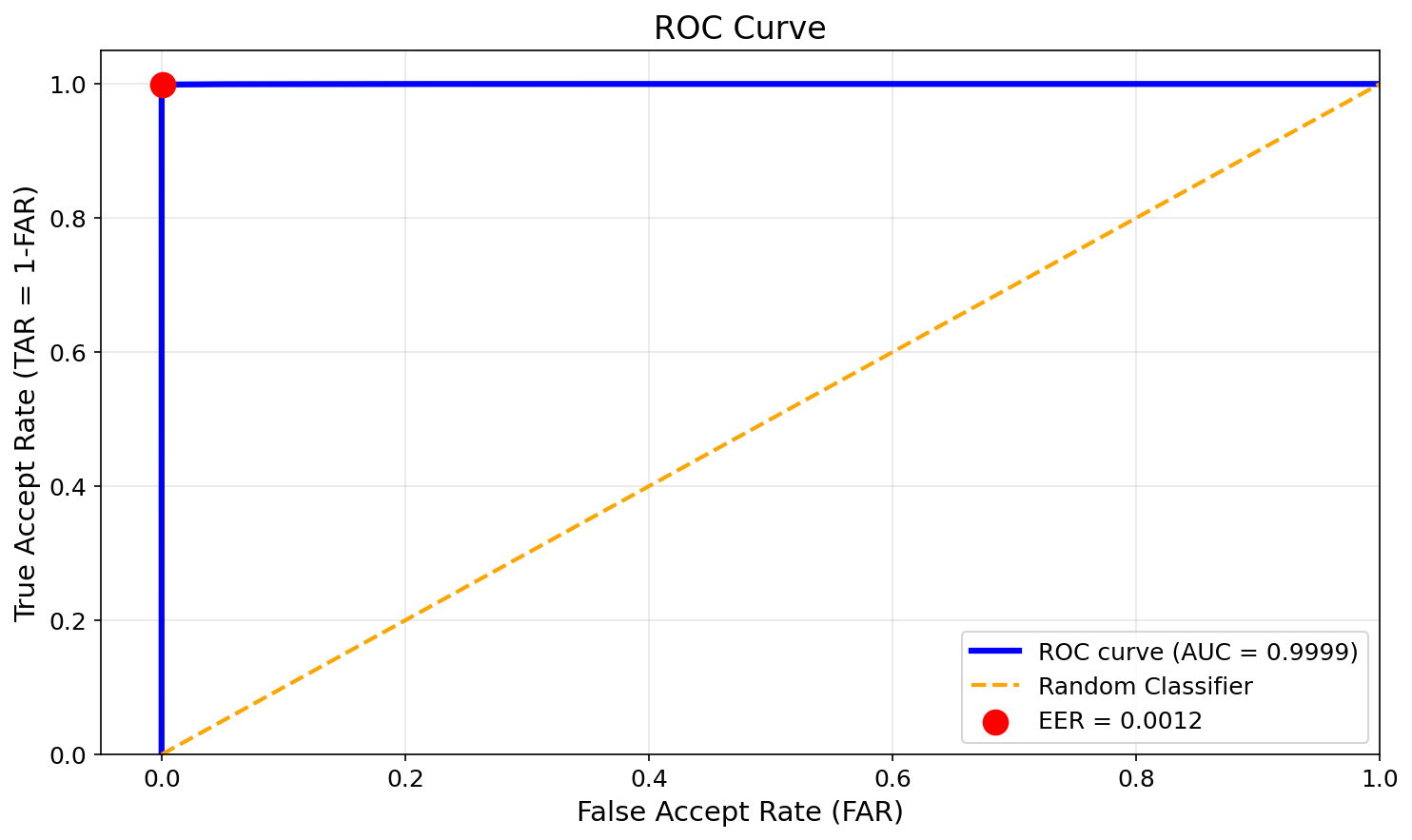}
        \caption{ROC Curve and EER}
        \label{fig:roc_eer_1K_sub}
    \end{subfigure}
    \begin{subfigure}[b]{0.46\linewidth}
        \includegraphics[width=\linewidth]{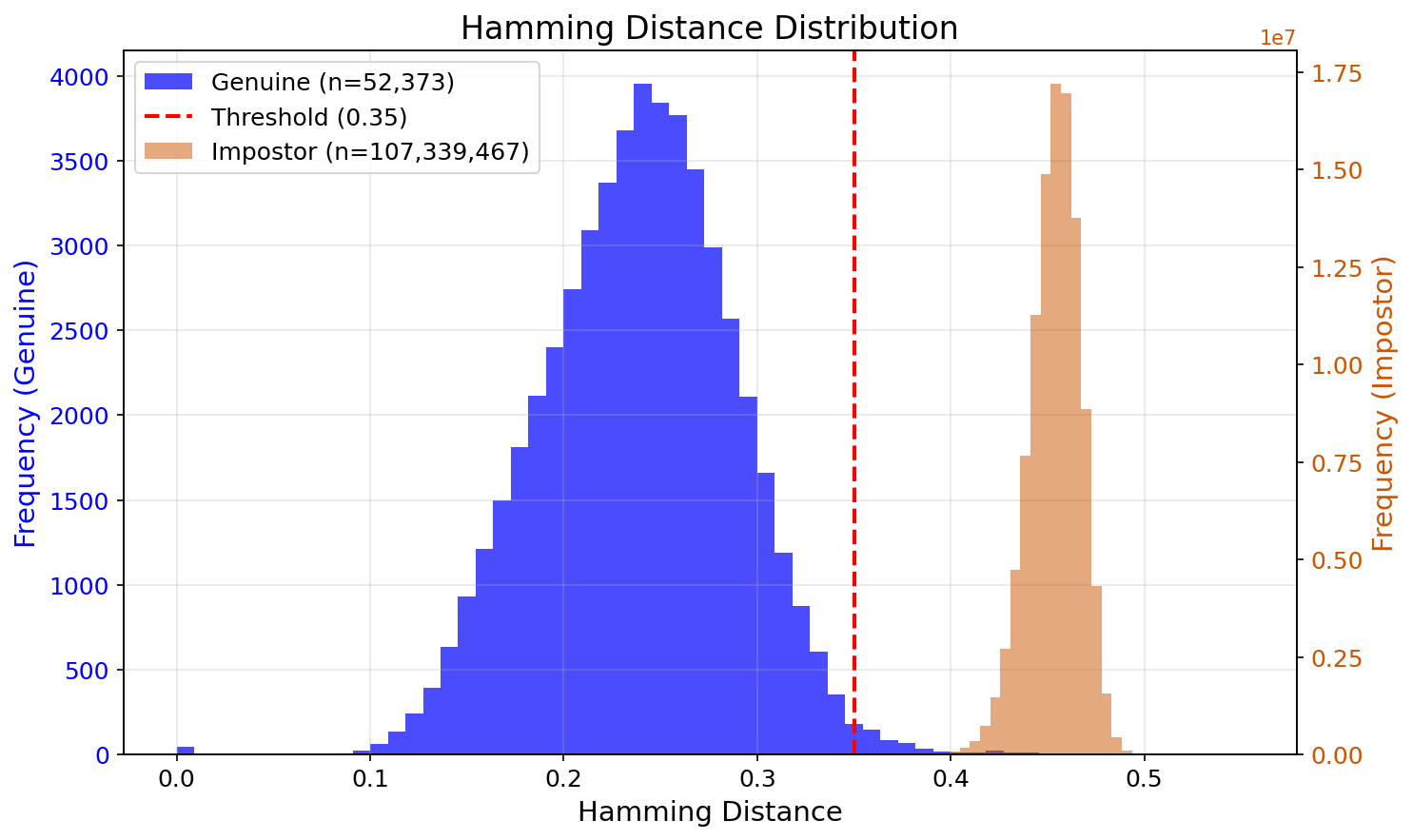}
        \caption{Hamming Distance Distributions}
        \label{fig:hd_distribution_1k_sub}
    \end{subfigure}
    \caption{Performance on the \textit{CASIA 1K} dataset. The EER is approximately 0.12\% at a threshold of 0.35, with an AUC of 0.9999.}
    \label{fig:combined_results}
    \vspace{-4mm} 
\end{figure}

The ROC curve and Hamming distance distributions for the largest 1\,000-subjects test (\textit{CASIA 1K}), shown in Figure~\ref{fig:combined_results}, illustrate the high separability between genuine and imposter populations ($d' \approx 5.96$). Unlike earlier baselines, the improved mean Impostor Hamming Distance ($\approx 0.45$) provides a significant safety margin against false matches. The ROC curves for the smaller datasets were highly similar and are provided in the Appendix~\ref{appendix:a2}.

\vspace{-0.75cm}
\begin{table}[h]
\centering
\caption{Performance Metrics Across Different Dataset Scales}
\label{tab:scalability_results}
% \vspace{-1mm} 
\begin{tabular}{@{}lcccc@{}}
\toprule
\textbf{Metric} & \textbf{CASIA 50} & \textbf{CASIA 100} & \textbf{CASIA 500} & \textbf{CASIA 1K} \\
\midrule
Total Comparisons & 300.7k & 1.19M & 24.9M & 107.4M \\
Genuine HD (Mean) & 0.242 & 0.243 & 0.241 & 0.239 \\
Impostor HD (Mean) & 0.454 & 0.454 & 0.454 & 0.454 \\
\textbf{EER (\%)} & \textbf{0.27} & \textbf{0.16} & \textbf{0.16} & \textbf{0.12} \\
\textbf{F1 Score} & \textbf{0.9945} & \textbf{0.9949} & \textbf{0.9928} & \textbf{0.9948} \\
\bottomrule
\end{tabular}
\vspace{-3mm} 
\end{table}

\vspace{-0.75cm}

\subsection{Homomorphic Encryption Performance}
For the FHE evaluation, \textit{1--1} comparisons on representative genuine (same eye, different images) and imposter (different subjects) pairs demonstrated high precision with a significant but predictable computational overhead. The complete processing pipeline for the genuine pair comparison, from segmentation to iris code generation, is visualized in Figure~\ref{fig:genuine_pair_pipeline}; a similar visualization for the imposter pair is provided in the Appendix \ref{appendix:a2}.

\begin{figure}[htbp]
    \centering
    \includegraphics[width=0.9\textwidth]{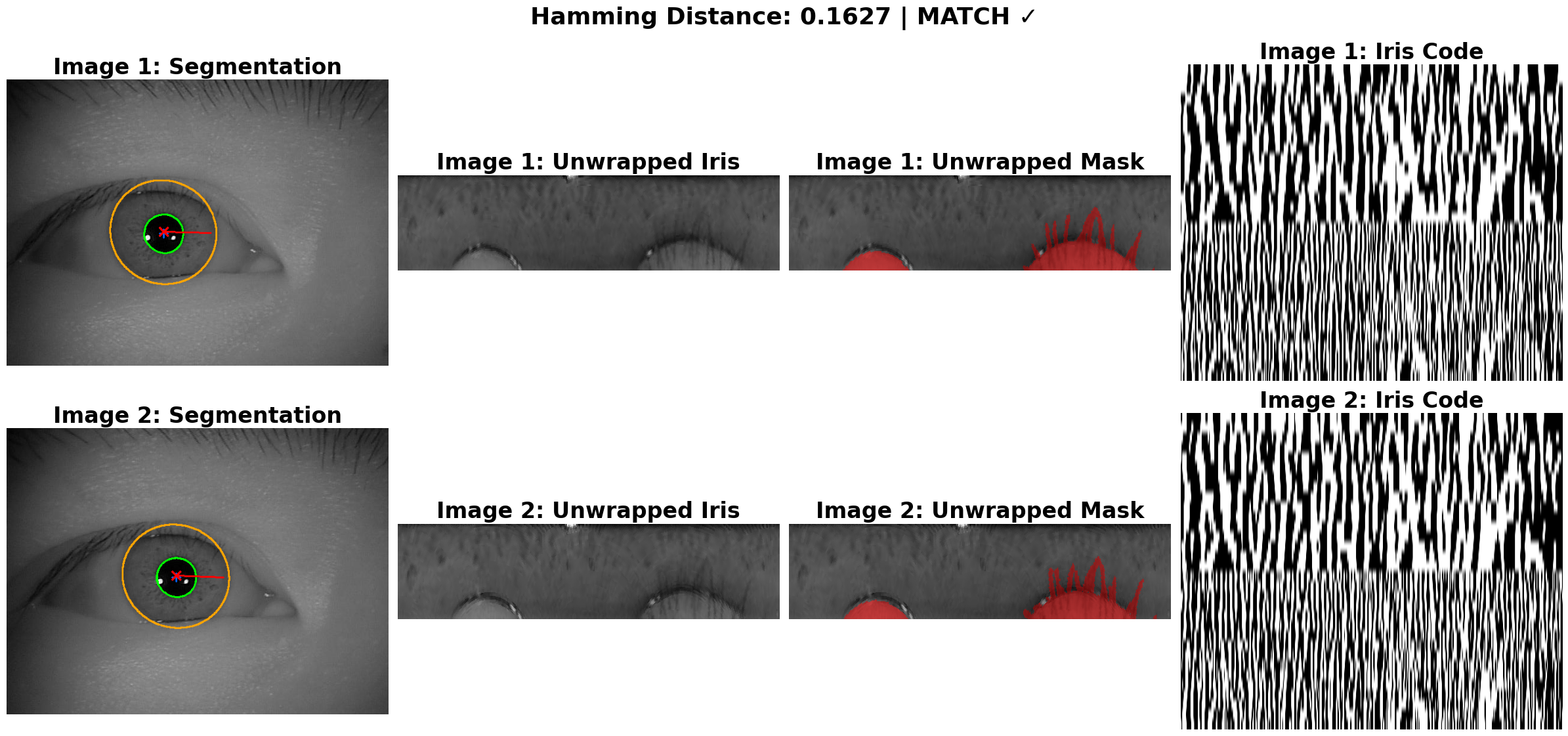}
    \caption{Processing and matching pipeline for a genuine pair (HD = 0.1627).}
    \label{fig:genuine_pair_pipeline}
    \vspace{-2mm}
\end{figure}

As detailed in Table~\ref{tab:he-vs-plain}, a single encrypted match took approximately 295--299 seconds, dominated by the homomorphic computation time ($\approx$ 296s), with fast encryption ($\approx$ 3.0s) and decryption ($<$ 0.01s). The decrypted Hamming distance for a genuine pair (0.1627) was identical to the cleartext result. For the imposter pair, the FHE result (0.4701) showed a negligible difference ($0.0002$) compared to the cleartext baseline (0.4703), confirming that our privacy-preserving framework maintains the accuracy of the underlying biometric algorithm.

\vspace{-0.75cm}
\begin{table}[htbp]
\centering
\caption{Cleartext vs. FHE Comparison for Genuine and Imposter Pairs}
\label{tab:he-vs-plain}
% \vspace{-1mm} 
\begin{tabular}{@{}llrrrr@{}}
\toprule
\textbf{Pair} & \textbf{Method} & \textbf{HD} & \textbf{Numerator} & \textbf{Denominator} & \textbf{Time (s)} \\
\midrule
\multirow{2}{*}{Genuine} & Cleartext & 0.1627 & 1\,914 & 11\,761 & 0.003 \\
                         & FHE       & 0.1627 & 1\,914 & 11\,761 & 299.4 \\
\midrule
\multirow{2}{*}{Imposter} & Cleartext & 0.4703 & 4\,736 & 10\,071 & 0.003 \\
                          & FHE       & 0.4701 & 4\,734 & 10\,071 & 295.1 \\
\bottomrule
\end{tabular}
\end{table}

\section{Discussion}
\label{sec:discussion}
The experimental results demonstrate the feasibility and reliability of iris recognition using homomorphic encryption-based matching. The cleartext baseline on the \textit{CASIA 1K} dataset yielded an Equal Error Rate (EER) of just 0.12\% and a separability index ($d'$) of 5.96, confirming that our segmentation, normalization, and encoding pipeline is highly effective even at scale. The low False Rejection Rate (FRR) of approx. 0.96\% suggests that the preprocessing steps handle common noise factors, such as eyelid occlusion and partial illumination, robustly.

Our proposed privacy-preserving pipeline achieves effectively identical matching accuracy under homomorphic encryption as in cleartext. For genuine pairs, the decrypted Hamming distance was identical to the plaintext baseline, while impostor pairs showed negligible deviation. This validates the mathematical correctness of our FHE implementation for biometric authentication. However, this security comes with a significant computational cost. A single \textit{1--1} matching operation, including encryption and 31 rotational shifts, requires approximately 295--300 seconds. While this latency is currently too high for real-time user verification, it proves the viability of the approach for high-security, asynchronous verification scenarios where privacy is paramount.

Applying an FHE scheme end-to-end enables strong recognition accuracy without exposing biometric templates. The primary bottleneck identified in our analysis is the homomorphic execution of rotational shifts to correct for head tilt, which accounted for over 98\% of the total computation time. These results are part of our ongoing work to balance security and usability. Next steps include exploring more sophisticated occlusion masks and analyzing the trade-offs of restricting the rotational search space to reduce latency.

 \vspace{-2mm} 
 \paragraph{\textbf{Limitations:}} While our results establish a strong proof-of-concept, the current implementation is not yet a production-ready solution for real-time applications. The latency introduced by the homomorphic operations, specifically the iterative calculation of Hamming distances across multiple rotations, remains the critical challenge. Future work will focus on optimizing this computational overhead. We aim to explore advanced FHE optimization techniques, such as batching multiple comparisons or utilizing schemes like CKKS, which could offer significant speedups by vectorizing the distance computation at the cost of negligible approximation.

\section{Conclusions and Future Work}
\label{sec:conclusions}
This paper investigated the key performance challenges of privacy-preserving iris recognition systems, demonstrating that Fully Homomorphic Encryption (FHE) can protect sensitive biometric data throughout the entire authentication pipeline without compromising accuracy. Our experiments on the CASIA-Iris-Thousand dataset confirmed that the cryptographic overhead does not degrade the underlying matching performance, achieving an Equal Error Rate (EER) of 0.12\% and effectively identical Hamming Distance calculations in the encrypted domain compared to cleartext. The framework meets critical privacy-by-design standards, making it suitable for high-security, decentralized identity applications. However, we confirmed that the primary barrier to real-time adoption remains the significant computational latency, particularly for the iterative task of secure rotational alignment.

\subsection{Future Work}
Our future research will focus on performance optimization and scalability to bridge the gap between proof-of-concept and production deployment. To enable scalable \textit{1--N} template comparisons, we will deploy a two-level hierarchical search strategy. This approach will utilize a fast, lightweight screening tier to rapidly filter out non-matching candidates, reserving the computationally intensive full FHE comparisons only for the most promising matches. To address the current latency bottlenecks, we will explore more efficient FHE schemes, such as CKKS, which can offer significant speedups by vectorizing Hamming distance computations through approximate arithmetic. To enhance practical utility for decentralized identity, we will work towards implementing essential features such as template revocability and secure updates. Through these directions, we aim to bridge the gap between theoretical cryptographic primitives and a scalable privacy-preserving biometric identification solution.


\section*{Appendix}
\label{appendix:a}
\appendix
\label{appendix:a}
\section{Supplementary Details and Figures}
This appendix provides detailed information on the experimental setup, a detailed breakdown of performance metrics at different operational thresholds for the largest dataset, and supplementary performance figures for the smaller datasets.

\subsection{Experimental Setup}
\label{appendix:a1}
\begin{itemize}
\item \textbf{Hardware:} All experiments were conducted on a single GNU/Linux system GPU machine equipped with an NVIDIA GeForce GTX 1\,080 graphics card and 32 GB of system memory.
\item \textbf{Software:} The iris processing pipeline, including segmentation and feature encoding, utilized the \textit{Open Iris} library~\cite{open-iris-repo}. The FHE implementation was built using Vaultree’s \textit{venumML}~\cite{vaultree2025} library. All cryptographic parameters were configured to provide a 128-bit security level, ensuring protection against all known classical and quantum computational attacks. 
\item \textbf{Dataset:} The evaluation was conducted using subsets of the publicly available CASIA-Iris-Thousand dataset, which contains 20\,000 iris images from 1\,000 subjects (10 images for each eye of 1\,000 subjects) \cite{CASIA-Iris}. The 640 × 480 pixel, 8-bit grayscale JPEG images feature clear, well-centered iris patterns, making them ideal for biometric identification. We removed the images with glasses, as they affect the image quality, and created four template databases of varying sizes to assess scalability and performance:
\begin{enumerate}
    \item \textbf{CASIA 50}: A database of 50 subjects containing 755 iris templates.
    \item \textbf{CASIA 100}: A database of 100 subjects containing 1\,487 iris templates.
    \item \textbf{CASIA 500}: A database of 500 subjects containing 6\,777 iris templates.
    \item \textbf{CASIA 1K}: A database of 1\,000 subjects containing 14\,191  iris templates.
\end{enumerate}
\end{itemize}

All possible intra-class (genuine) and inter-class (imposter) comparisons were performed on these databases to generate Hamming distance (HD) distributions and evaluate the system's discriminative power. For the FHE evaluation, we performed \textit{1--1} comparisons on representative genuine and imposter pairs to precisely measure computational overhead and accuracy.

\subsection{Supplementary Figures}
\label{appendix:a2}

% CASIA 50
\begin{figure}[h!]
    \centering
    % Left: Distribution
    \begin{subfigure}[b]{0.49\linewidth}
        \includegraphics[width=\linewidth]{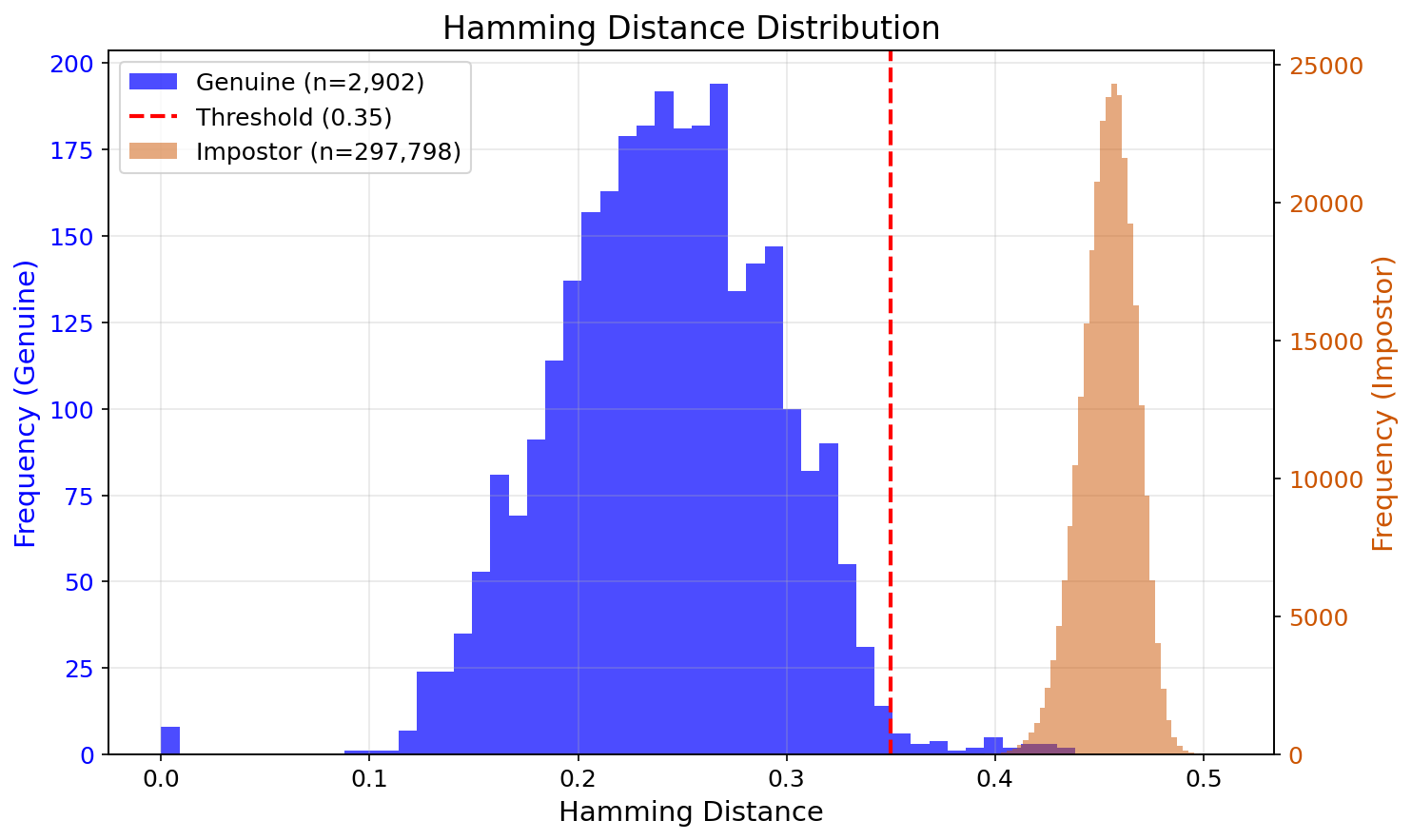}
        \caption{Hamming Distance Distribution}
        \label{fig:hd_dist_50_sub}
    \end{subfigure}
    \hfill
    % Right: ROC
    \begin{subfigure}[b]{0.49\linewidth}
        \includegraphics[width=\linewidth]{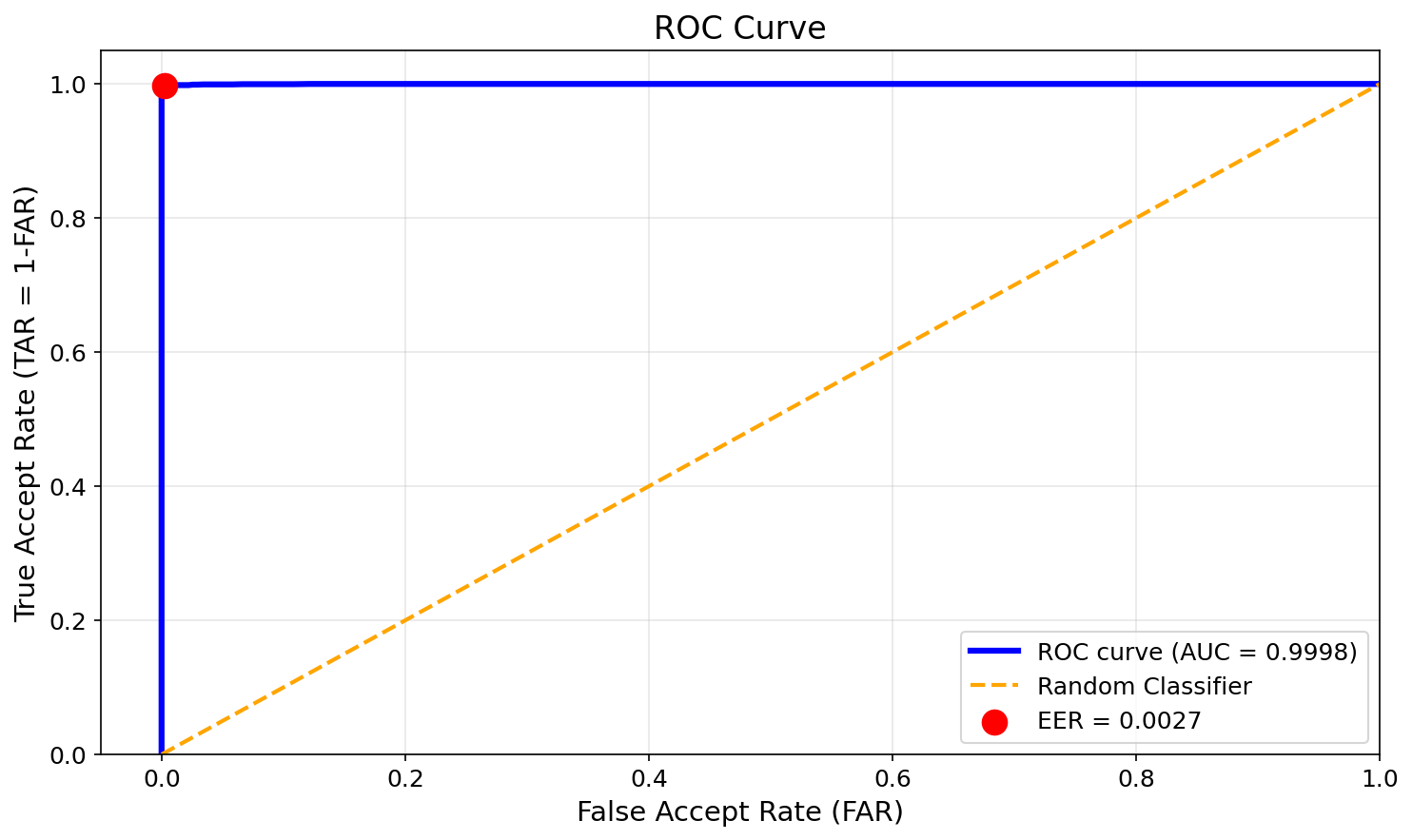}
        \caption{ROC Curve \\ \tiny{($EER=0.27\%, AUC=0.9999$)}}
        \label{fig:roc_eer_50_sub}
    \end{subfigure}
    \caption{Hamming distance distributions and ROC curve for \textit{CASIA 50}.}
    \label{fig:casia_50_eval}
\end{figure}

% CASIA 100
\begin{figure}[h!]
    \centering
    % Left: Distribution
    \begin{subfigure}[b]{0.49\linewidth}
        \includegraphics[width=\linewidth]{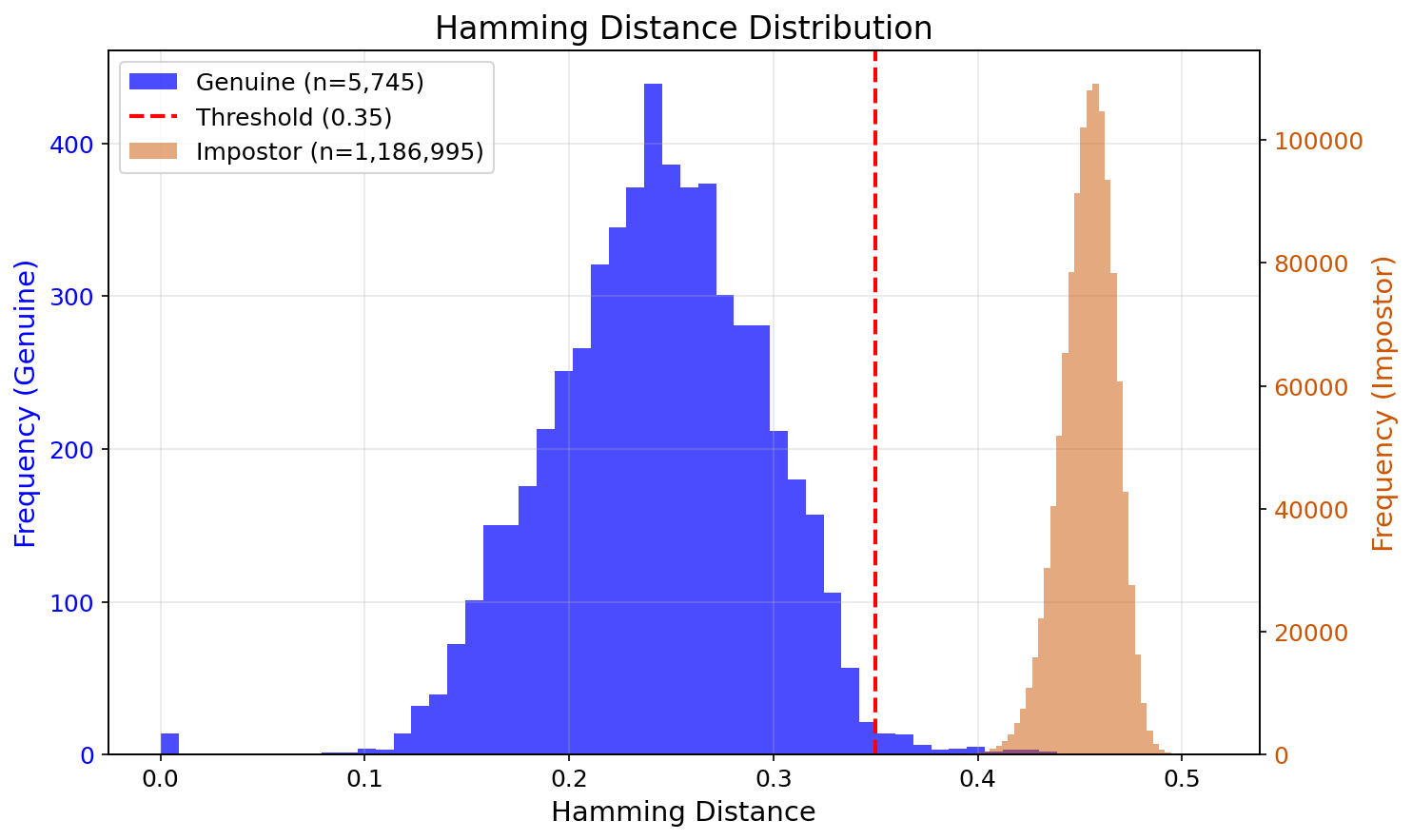}
        \caption{Hamming Distance Distribution}
        \label{fig:hd_dist_100_sub}
    \end{subfigure}
    \hfill
    % Right: ROC
    \begin{subfigure}[b]{0.49\linewidth}
        \includegraphics[width=\linewidth]{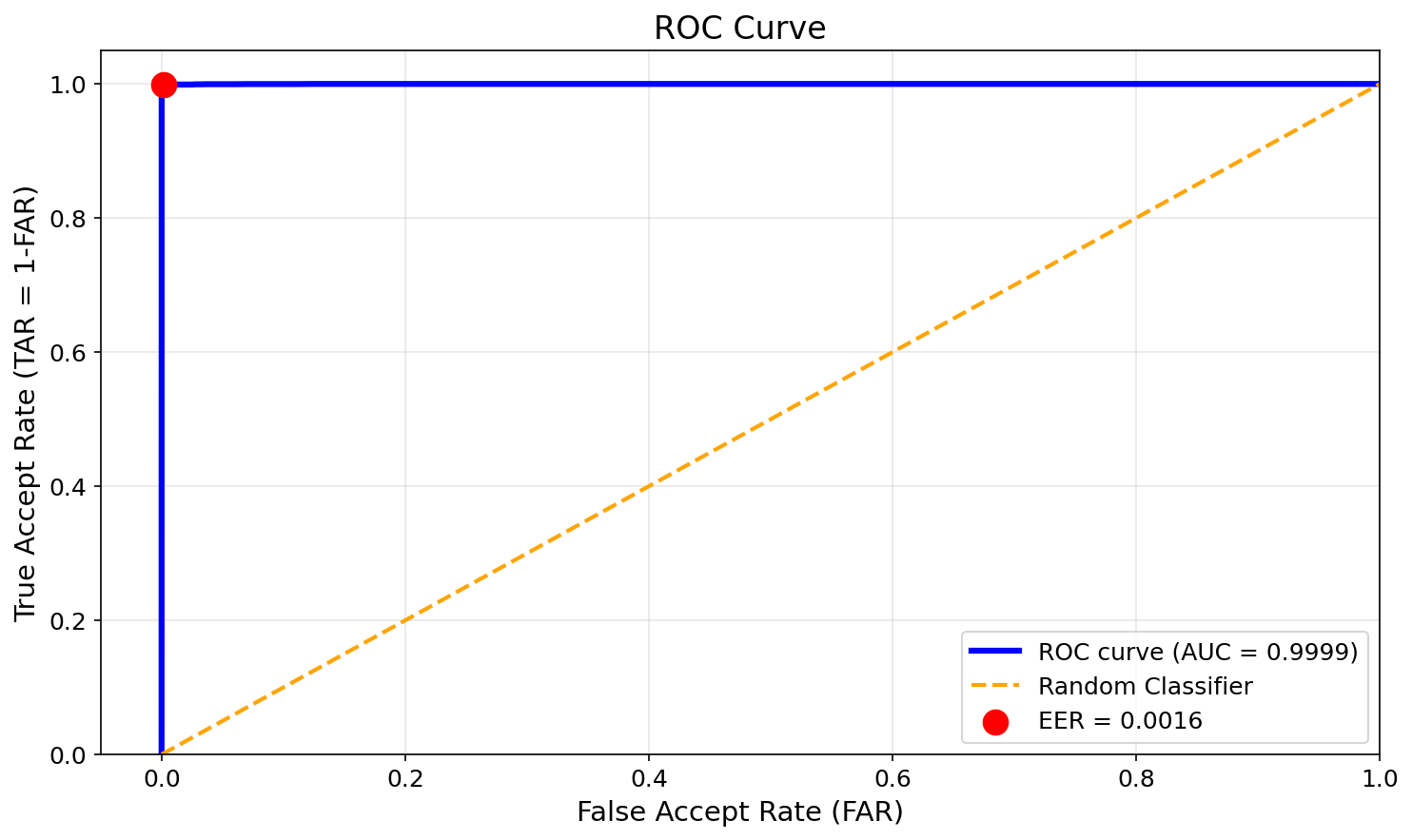}
        \caption{ROC Curve \\ \tiny{($EER=0.16\%, AUC=0.9999$)}}
        \label{fig:roc_eer_100_sub}
    \end{subfigure}
    \caption{Hamming distance distributions and ROC curve \textit{CASIA 100}.}
    \label{fig:casia_100_eval}
\end{figure}

% CASIA 500
\begin{figure}[h!]
    \centering
    % Left: Distribution
    \begin{subfigure}[b]{0.49\linewidth}
        \includegraphics[width=\linewidth]{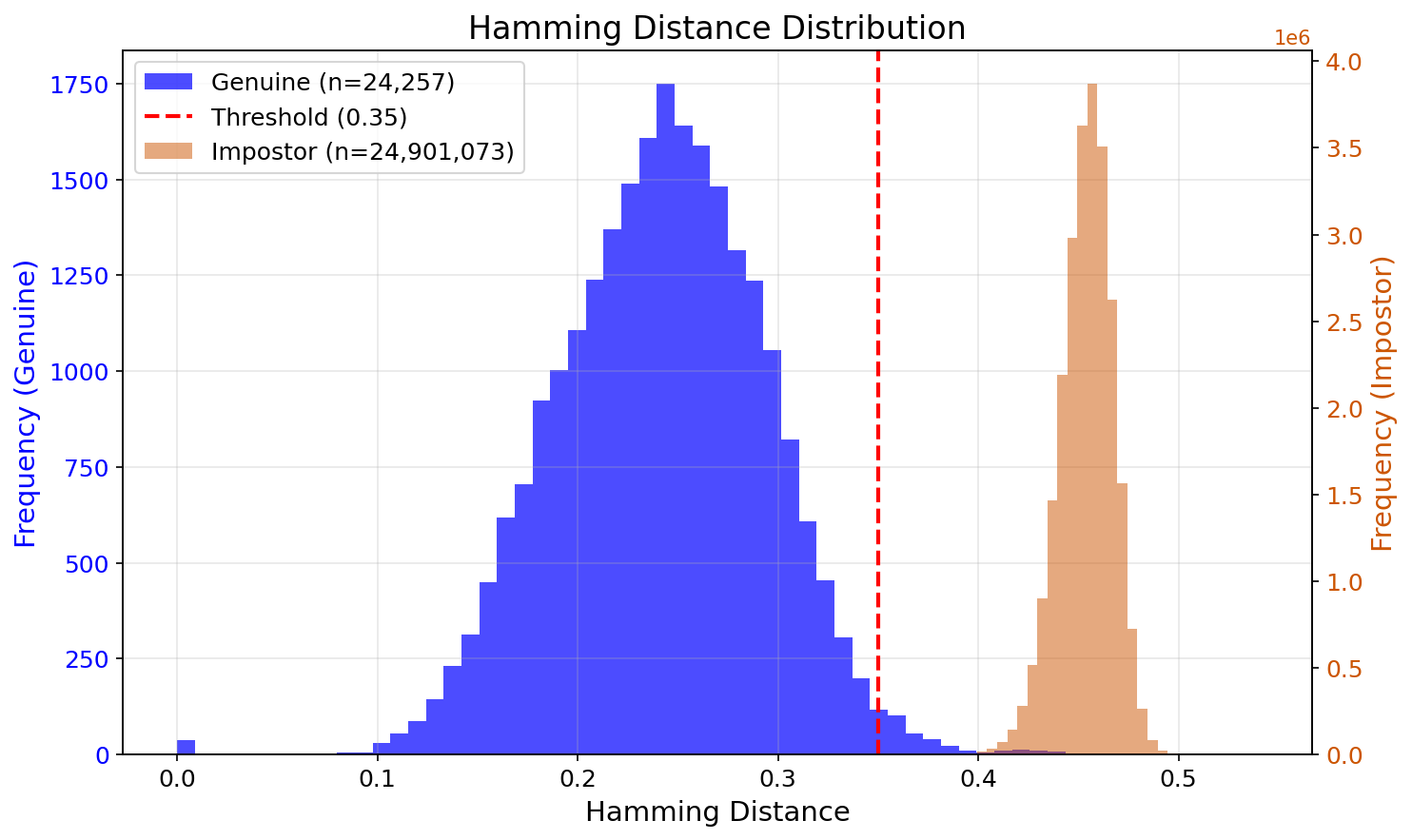}
        \caption{Hamming Distance Distribution}
        \label{fig:hd_dist_500_sub}
    \end{subfigure}
    \hfill
    % Right: ROC
    \begin{subfigure}[b]{0.49\linewidth}
        \includegraphics[width=\linewidth]{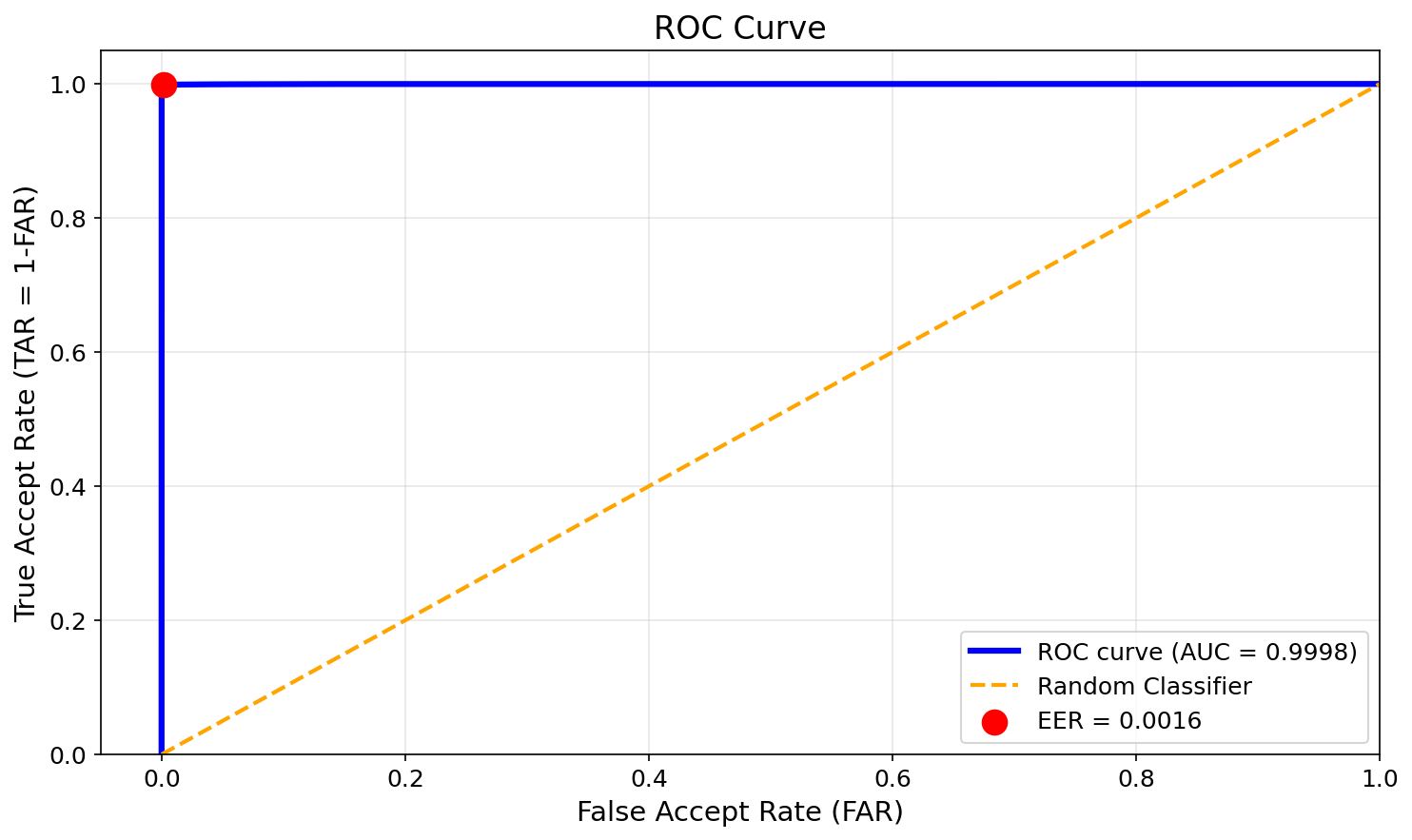}
        \caption{ROC Curve \\ \tiny{($EER=0.16\%, AUC=0.9999$)}}
        \label{fig:roc_eer_500_sub}
    \end{subfigure}
    \caption{Hamming distance distributions and ROC curve \textit{CASIA 500}.}
    \label{fig:casia_500_eval}
\end{figure}

\begin{figure}[h!]
    \centering
    \includegraphics[width=0.9\textwidth]{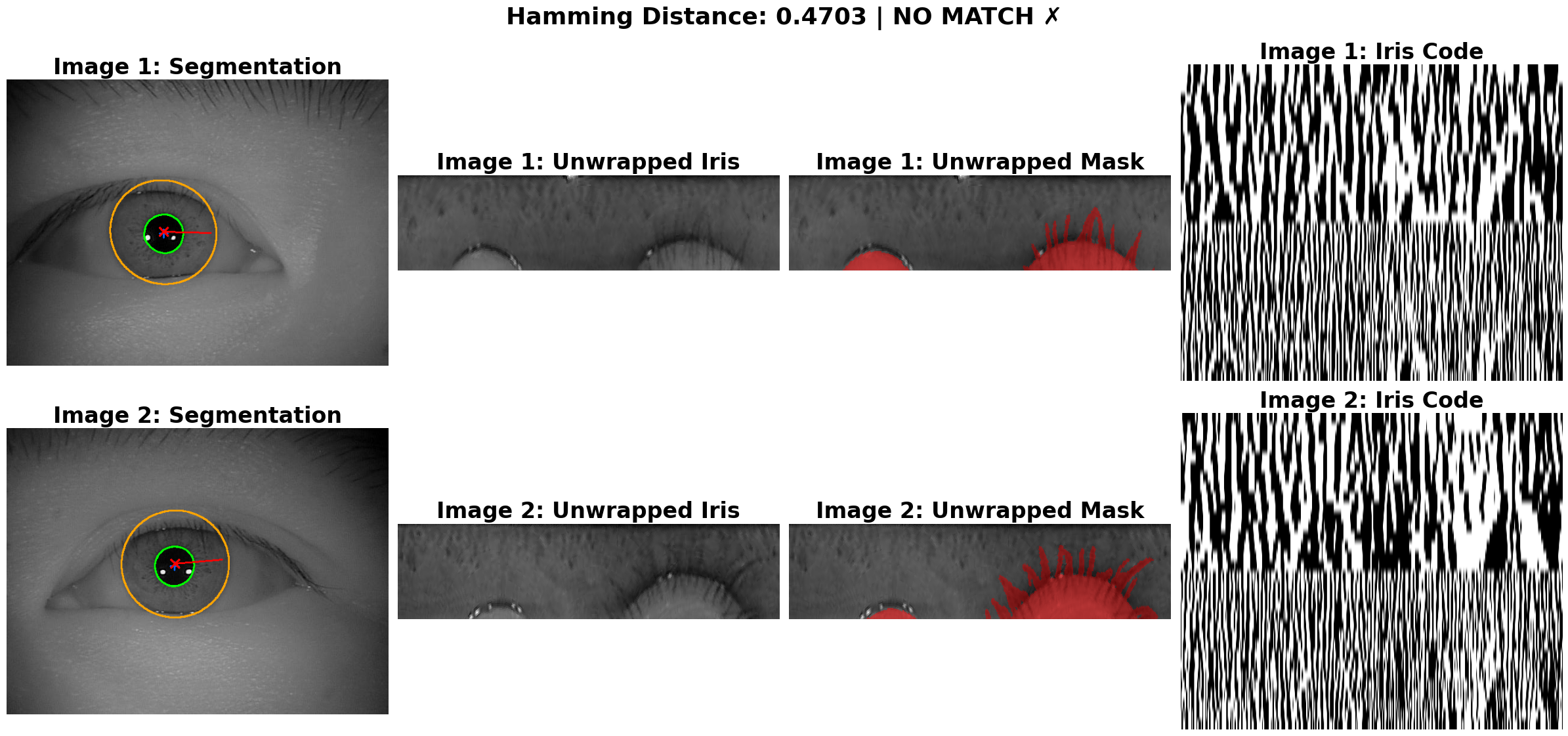}
    \caption{Visualisation of the full processing pipeline for an imposter pair comparison, resulting in a non-match Hamming distance of 0.4703.}
    %The resulting Hamming distance of 0.4355 is correctly classified as a non-match.
    \label{fig:impostor_pair_pipeline}
\end{figure}

\end{document}